\documentclass[aps,amsmath, prb,twocolumn,superscriptaddress]{revtex4-2}
\usepackage{graphicx}
\usepackage{dcolumn}
\usepackage{bm}
\usepackage{lipsum}

\begin{document}

\preprint{}
\title{Strain-induced Shape Anisotropy in Antiferromagnetic Structures}

\author{Hendrik Meer}
\email[]{meer@uni-mainz.de}
\affiliation{Institute of Physics, Johannes Gutenberg-University Mainz, 55099 Mainz, Germany}

\author{Olena Gomonay}
\affiliation{Institute of Physics, Johannes Gutenberg-University Mainz, 55099 Mainz, Germany}
\author{Christin Schmitt}
\affiliation{Institute of Physics, Johannes Gutenberg-University Mainz, 55099 Mainz, Germany}

\author{Rafael Ramos}
\affiliation{WPI-Advanced Institute for Materials Research, Tohoku University, Sendai 980-8577, Japan}
\affiliation{Centro de Investigación en Química Biolóxica e Materiais Moleculares (CIQUS), Departamento de Química-Física, Universidade de Santiago de Compostela, Santiago de Compostela 15782, Spain}

\author{Leo Schnitzspan}
\affiliation{Institute of Physics, Johannes Gutenberg-University Mainz, 55099 Mainz, Germany}

\author{Florian Kronast}
\affiliation{Helmholtz-Zentrum Berlin für Materialien und Energie, Albert-Einstein-Strasse 15, 12489 Berlin, Germany}

\author{Mohamad-Assaad Mawass}
\affiliation{Helmholtz-Zentrum Berlin für Materialien und Energie, Albert-Einstein-Strasse 15, 12489 Berlin, Germany}

\author{Sergio Valencia}
\affiliation{Helmholtz-Zentrum Berlin für Materialien und Energie, Albert-Einstein-Strasse 15, 12489 Berlin, Germany}

\author{Eiji Saitoh}

\affiliation{WPI-Advanced Institute for Materials Research, Tohoku University, Sendai 980-8577, Japan}
\affiliation{Department of Applied Physics, The University of Tokyo, Tokyo 113-8656, Japan}
\affiliation{Center for Spintronics Research Network, Tohoku University, Sendai 980-8577, Japan}
\affiliation{Advanced Science Research Center, Japan Atomic Energy Agency, Tokai 319-1195, Japan}

\author{Jairo Sinova}
\affiliation{Institute of Physics, Johannes Gutenberg-University Mainz, 55099 Mainz, Germany}
\affiliation{Institut of Physics, Academy of Sciences of the Czech Republic, Praha 11720, Czech Republic}

\author{Lorenzo Baldrati}
\affiliation{Institute of Physics, Johannes Gutenberg-University Mainz, 55099 Mainz, Germany}

\author{Mathias Kl\"aui}
\email[]{Klaeui@Uni-Mainz.de}
\affiliation{Institute of Physics, Johannes Gutenberg-University Mainz, 55099 Mainz, Germany}
\affiliation{Center for Quantum Spintronics, Department of Physics, Norwegian University of Science and Technology,
7034 Trondheim, Norway}

\date{\today}

\begin{abstract}
We demonstrate how shape-dependent strain can be used to control antiferromagnetic order in NiO/Pt thin films. For rectangular elements patterned along the easy and hard magnetocrystalline anisotropy axes of our film, we observe different domain structures and we identify magnetoelastic interactions that are distinct for different domain configurations. We reproduce the experimental observations by modeling the magnetoelastic interactions, considering spontaneous strain induced by the domain configuration, as well as elastic strain due to the substrate and the shape of the patterns. This allows us to demonstrate and explain how the variation of the aspect ratio of rectangular elements can be used to control the antiferromagnetic ground state domain configuration. Shape-dependent strain does not only need to be considered in the design of antiferromagnetic devices, but can potentially be used to tailor their properties, providing an additional handle to control antiferromagnets.
\end{abstract}
\maketitle
\newpage

Antiferromagnetic materials (AFM) are promising candidates for fast, robust and energy-efficient spintronic devices \cite{Baltz2018}. AFMs possess two or more magnetic sublattices, with vanishing net magnetization. This absence of magnetic stray fields enables higher bit packing density of AFMs devices compared to ferro(i)magnetic materials (FMs), enhanced robustness against interfering external magnetic fields and potentially THz switching speeds \citep{Kampfrath2011}. Especially insulating antiferromagnetic materials (iAFM) have emerged as a promising material class for the development of low power devices, because their low damping allows for the transport of spin currents over long distances \citep{Lebrun2018}.

Crucial for the implementation of AFMs as active spintronic devices is the control of the antiferromagnetic order. In recent years it has been established that current pulses through an adjacent heavy metal layer can induce a reorientation of the antiferromagnetic ordering in insulating AFMs \citep{Moriyama2018, Chen2018, Baldrati2019}. For iAFMs with strong magnetostriction, the reorientation of the N\'eel  vector is dominated by a thermomagnetoelastic switching and strongly depends on the device geometry \citep{Zhang2018, Baldrati2020, Meer2021}.

For FMs the device geometry and shape-induced control of the domains is a key tool for tailoring functional device properties. In AFMs, conventional shape anisotropy caused by the magnetic dipolar interactions is not present, due to the absence of a demagnetization field. However, theoretical work on shape-induced phenomena in finite size antiferromagnets predicts an ordering of the antiferromagnetic domains, with long-range strain fields leading to the formation of shape-dependent domain structures \citep{Gomonay2007, Gomonay2014}. It has been shown that the shape-induced domains in the FM layer of a AFM/FM bilayer can for instance be used to imprint an AFM vortex state into the adjacent AFM layer \citep{Wu2011, Sort2006}. However, initial studies of patterning-induced effects in antiferromagnetic LaFeO$_3$ could not observe any changes in the domain structure, after patterning different elements with etching \citep{Czekaj2007}. Later studies patterned elements via an Ar$^+$ ion implantation-based patterning technique, which resulted in antiferromagnetic structures embedded in a non-magnetic layer \citep{Folven2010, Lee2020}. This technique led to the observation of changes in the antiferromagnetic ordering near the patterning edge for LaFeO$_3$ \citep{Folven2010, Folven2011, Folven2012} and more recently La$_{0.7}$Sr$_{0.3}$FeO$_3$ \citep{Lee2020}, interpreted as an edge effect near the edge for elements patterned along the easy axis.
Further studies have suggested the exploitation of this effect in exchange bias applications of AFM/FM heterostructures \citep{Folven2015}.  However, these previous investigations of patterning-induced modulations of the AFM order have been focused on the passive application of AFMs in AFM/FM bilayers. Considering the potential of antiferromagnets as active elements in spintronic devices, it is important to investigate patterning- and shape-induced effects and in particular the control of the domain configuration in AFMs without an adjacent FM layer. It is crucial to not only understand patterning-induced effects near the edge, but also the influence of  shape-dependent strain on the domain structure inside a structured antiferromagnetic device. This effect would be most suitable to tailor domain configurations by the shape.

The prototypical collinear antiferromagnet NiO has been considered to be a promising candidate for an active element in spintronic applications, in contrast to LaFeO$_3$, due to the possibility of electrically controlling and reading the AFM order \citep{Hoogeboom2017, Moriyama2018, Baldrati2018a, Fischer2018} and recent observations of ultrafast currents in the THz regime in NiO/Pt bilayers \citep{Kampfrath2011, Moriyama2020}. In addition, NiO exhibits a high N\'eel  temperature of 523$\,$K in the bulk \citep{Roth1960} and strong magneto-elastic coupling \citep{Aytan2017}. The latter has been used extensively to manipulate the AFM order of NiO by growth-induced strain \citep{Kozio-Rachwa2020, Altieri2003, Alders1998}, piezoelectric substrates exerting strain \citep{Barra2021} and indirectly via current-induced heating leading to strain \citep{Meer2021}. However, the effect of shape-dependent strain on the domain structure of NiO thin films has not been explored. Considering the application of AFMs with strong magnetostriction like NiO or CoO in active spintronic devices, it is important to investigate how the geometry influences the antiferromagnetic domain configuration and how one could use different geometries to control the antiferromagnetic order.

In this work, we demonstrate the tailoring of the AFM ground state domain configuration of NiO by shape-dependent strain. We study the N\'eel vector orientation in patterned elements by photoemission electron microscopy (PEEM) exploiting the x-ray magnetic linear dichroism (XMLD) effect for magnetic contrast. We first identify and compare the shape-induced domain structure of elements oriented along different axes before we theoretically explore how shape-induced effects can manipulate the antiferromagnetic ordering in different element geometries. Finally, we demonstrate how the modification of the shape-dependent strain by variation of the aspect ratio of our elements can be used to control the antiferromagnetic domain configuration, demonstrating thus a tool for the shape-induced control of future AFM devices.


\section{Results} 
\begin{figure}[b]
\includegraphics{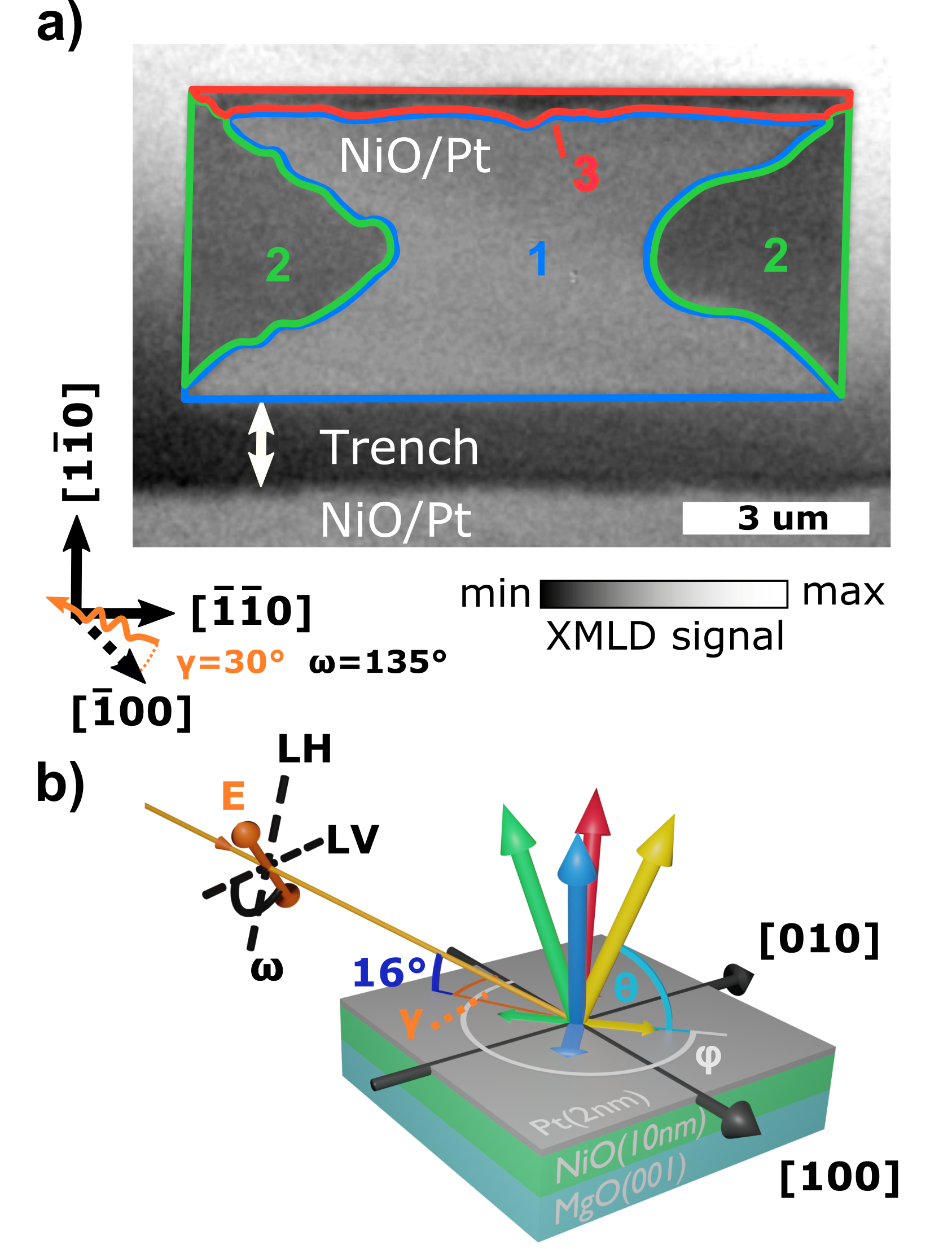}
\caption{\label{fig:1} (a) XMLD-PEEM image of shape-induced NiO domains inside a rectangular shaped element, with its edges oriented along the in-plane projection of the easy axis.  (b) Sketch of the experimental setup and orientation of the observed different N\'eel vector directions $[ \pm 5  \pm 5$ $19]$ with respect to the polarization vector. An orientation along $[5$ $5$ $19]$ (yellow) is equally possible, but was not observed in this particular element.}
\end{figure}

To investigate shape-induced effects on the antiferromagnetic domain structure, we have grown an epitaxial NiO(10nm)/Pt(2nm) bilayer on an MgO(001) substrate and used Ar ion beam etching to pattern various elements with different orientations. Similarly prepared bilayers of NiO and Pt are currently extensively used for current-induced switching \citep{Moriyama2018, Chen2018, Baldrati2019, Zhang2018, Baldrati2020, Meer2021} and THz radiation experiments \citep{Paperfrom2020}. As depicted in Fig.\ref{fig:1}a, we have etched trenches with a width of around $1\,$\textmu m and a depth of about $20\,$nm around the desired elements. Additionally, we deposited about $1.4\,$nm of ruthenium inside the trenches to reduce the possibility of discharges during PEEM imaging \citep{Stohr1999}. To allow for a reconfiguration of the AFM domains, the sample was annealed after pattering above its N\'eel temperature for 10 minutes at 550$\,$K under vacuum. Measurements were carried out at the UE49-PGM/SPEEM  at the BESSY II electron storage ring operated by the Helmholtz-Zentrum Berlin für Materialien und Energie \citep{Kronast2016}. \\

We first measured polarization-dependent absorption spectra around the Ni \textit{L}$_3$ and \textit{L}$_2$ edge (details of the measurement technique, see Appendix \ref{sec_XMLDsignal}) to verify the antiferromagnetic order of our films at room temperature. The XMLD contrast is proportional to the orientation of the N\'eel vector. By studying the XMLD contrast dependence on the azimuthal angle $\gamma$ and angle of the beam-polarization $\omega$ (Appendix \ref{sec_XMLDsignal}) we can identify in Fig. \ref{fig:1}a four antiferromagnetic domains present in a rectangular element of $10$\,\textmu m $\times5\,$\textmu m whose long axis is oriented along the [110] direction. Three different levels of XMLD contrast are observed inside our element, indicating three types of domains, a larger domain in the center (domain 1 - blue), two originating from the short edges (domain 2 - green) and one narrow at the long edge (domain 3 - red). The directions of the N\'eel  vector in the different domains are depicted in Fig. \ref{fig:1}b. We can observe that the in-plane projection of the N\'eel vector in all domains is oriented orthogonal to the edges of the element along [110] and $[1\bar{1}0]$.
The formation of domain 3 (red) along the edge can be attributed to localized changes of the anisotropy near the edge of the element, related to patterning-induced material property changes. However, the shape of domains 2 (green) and 1 (blue) in the center of the element can not be understood by local changes of the anisotropy at the edge of the element and we need to model long-range magnetoelastic interactions to understand the domain configuration.

NiO is known for its strong magnetoelastic coupling which is responsible for the creation of internal (magnetoelastic) stresses in a magnetically ordered state. In a free-standing homogeneously magnetically ordered sample, these stresses induce a pronounced spontaneous strain ($u_0\propto 10^{-4}- 3\cdot 10^{-3}$ \cite{Nakahigashi1975}) characterized by the strain tensor $\hat{u}^\mathrm{spon}_0$ whose components are related with the components of the N\'eel vector $\mathbf{n}$. 
In a multilayer system the internal magnetoelastic stresses are complemented by the external stresses due to the clamping of the antiferromagnetic layer by a nonmagnetic substrate \citep{Gomonay2002}. In this case the resulting strain field can be split into two parts, spontaneous (or plastic) strains $\hat{u}^\mathrm{spon}_0[\mathbf{n}(\mathbf{r})]$ associated with the distribution of the N\'eel vector (as in the absence of a substrate) and additional, elastic strains $\hat{u}^\mathrm{elast}$: $\hat{u}^\mathrm{tot}=\hat{u}^\mathrm{spon}_0+\hat{u}^\mathrm{elast}$.

To calculate the elastic strain, we use an approach of elasticity theory with continuously distributed defects \cite{Teodosiu1982, Kleman1972}. In particular, we assume that in magnetic multilayers the defects originate from the incompatibility between the spontaneous strain $\hat{u}^\mathrm{spon}_0$ and the non-deformed (reference) state of a non-magnetic substrate at the NiO/substrate interface, or from incompatibility between spontaneous strain in neighboring domains. From the compatibility condition for the total strain $\varepsilon_{ijk}\varepsilon_{lmn}\partial_j\partial_m{u}^\mathrm{tot}_{kn}  =0$ (where $\varepsilon_{ijk}$ is an antisymmetric Levi-Civita tensor) we obtain a set of equations for the elastic strains
\begin{equation}\label{eq_incompatibility}
\varepsilon_{ijk}\varepsilon_{lmn}\partial_j\partial_m{u}^\mathrm{elast}_{kn}  =\eta_{il},
\end{equation}
in which the incompatibility tensor ${\eta}_{il}\equiv-\varepsilon_{ijk}\varepsilon_{lmn}\partial_j\partial_m{u}^\mathrm{spon}_{kn}[\mathbf{n}(\mathbf{r})]$ is calculated for a given distribution of the N\'eel vector.

Equation (\ref{eq_incompatibility}) is similar to equations of electrostatics, in which the incompatibility $\hat{\eta}$ plays the role of the elastic or magnetoelastic charges and the elastic strains $\hat{u}^\mathrm{elast}$ correspond to the potentials \cite{Eshelby1956}. Moreover, similar to the electric and magnetostatic stray fields, the field of the corresponding elastic strains is long-range and therefore can stabilise an inhomogeneous distribution of the magnetic vectors. The similarity with the equations of electrostatics and magnetostatics allows for a qualitative interpretation of the magnetoelastic effects in terms of magnetoelastic charge distributions.

Here, we consider some of the effects that reinforce our intuitive reasoning through modelling.
The theoretical description of magnetic textures is based on minimizing the total energy of a sample with respect to magnetic and elastic variables. The bulk energy of the NiO film,
 \begin{equation}\label{eq_general_energy}
     W_\mathrm{bulk}=\int d\mathbf{r} (w_\mathrm{mag}+w_\mathrm{m-e}+w_\mathrm{elas})
 \end{equation}
includes magnetic, $w_\mathrm{mag}$, magnetoelastic, $w_\mathrm{m-e}$, and elastic, $w_\mathrm{elas}$ contributions. The magnetic structure of NiO is described by the N\'eel vector $\mathbf{n}(\mathbf{r})$ ($|\mathbf{n}|=1$), which is generally parameterized with two angles, $\varphi$  and $\vartheta$, as shown in Fig. 1b.

We distinguish in our thin NiO films four types of T-domains with the N\'eel vector oriented along $[ \pm 5  \pm 5$ $19]$ \cite{Schmitt2020b}. The pairs with opposite orientation of projection on the film plane have the same in-plane components of spontaneous strain and will be treated in the further discussion as the same domain. Since the out-of-plane component of the N\'eel vector is the same in all domains, we consider the in-plane angle $\varphi$ as the only magnetic variable. We also neglect the magnetic  homogeneity throughout the thickness of the NiO  layer and consider the distribution of the N\'eel vector to be within the film plane ($xy$).

We start with a discussion of the origin of the domain structure in thin films and patterned elements. A single-domain continuous film of NiO is charged due to incompatibility strain charges homogeneously distributed at the interface with the non-magnetic substrate. The charge density depends on the elastic and magnetoelastic properties of the interface and is localised in a thin layer of the order of the exchange length (characteristic length scale at which the N\'eel vector decays inside the nonmagnetic region), see Appendix~\ref{sec_theory}. These magnetoelastic charges create additional homogeneous strain $\hat{u}^\mathrm{elast}$  and their non-negative contribution to the energy of the NiO layer is proportional to the volume of the NiO.

\begin{widetext}

\begin{figure}[h]
\includegraphics{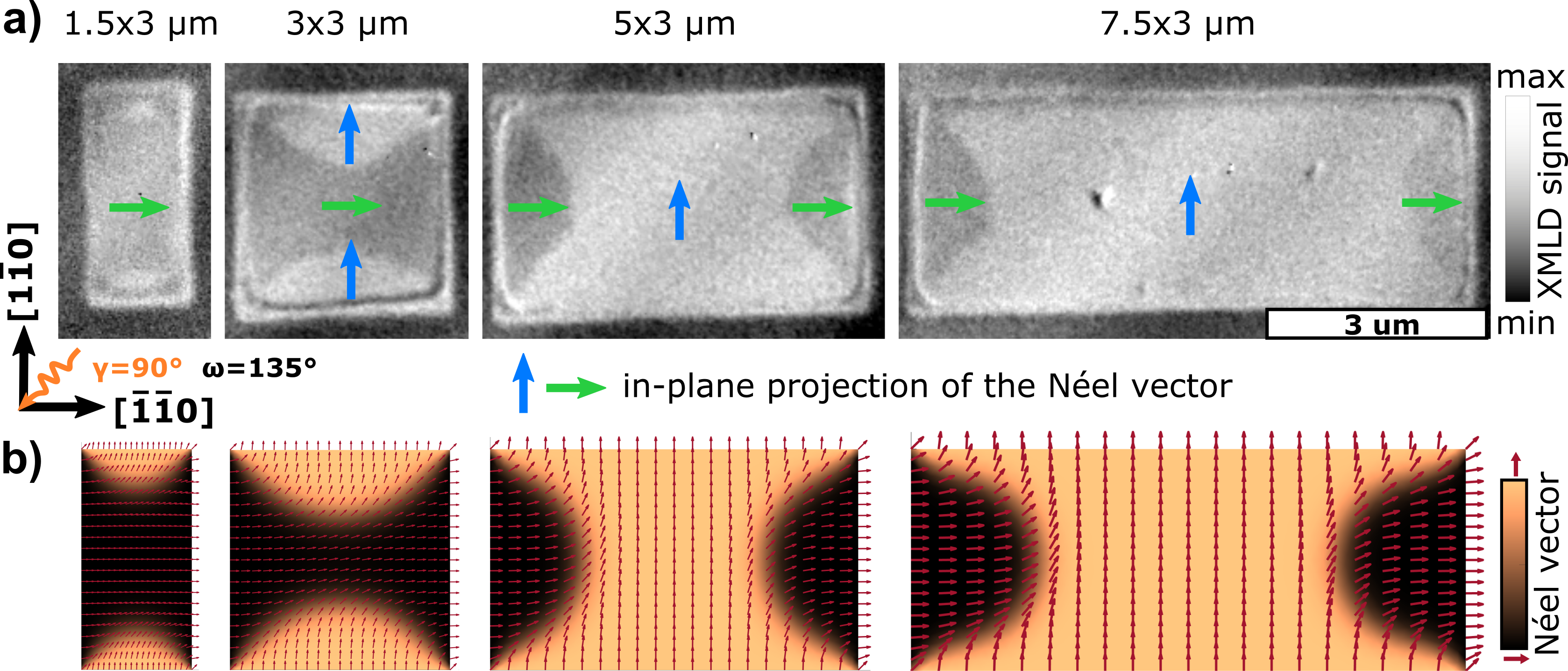}
\caption{\label{fig:3} (a) XMLD images of different NiO(10)/Pt(2) rectangular devices with varying aspect ratio. The edges of the devices are oriented along the in-plane projection of the easy axes. The arrows show the in-plane projection of the N\'eel vector determined from the greyscale contrast. (b) Final equilibrium state of the magnetic texture after considering magneto-elastic interactions to simulate the domain distribution in different aspect ratios. The color code indicates the direction of the N\'eel vector.}
\end{figure}

\end{widetext}

In a multidomain sample with equally distributed domains of all types, the average strain incompatibility and average charge density vanish. The local charge density is still non-zero and contributes to the energy of the sample. However, this contribution is proportional to the average domain volume. Hence, a small-scale multidomain structure is energetically favourable, with the domain size being limited by the positive energy contribution of the domain walls (similar to the Kittel model in ferromagnets \citep{Kittel1949}). However, the formation of a new domain inside a single-domain region is blocked by a high energy barrier associated with the coherent rotation of a large number of magnetic moments in the two sublattices. The energy barrier can be much lower at the sample surfaces and edges due to the additional contributions from surface energy and incompatibility charges at the element corners \citep{Gomonay2002}.

First, we consider the role of the surface magnetic anisotropy present in thin film NiO continuous films and patterned elements, which in our case favours alignment of the N\'eel vector perpendicular to the surface. For this we studied the evolution of the magnetic structure in the patterned elements with different aspect ratio cut parallel to the in-plane projection of the easy magnetic axis (see Fig. \ref{fig:3}a).

\begin{figure}
\includegraphics[width=0.5\textwidth]{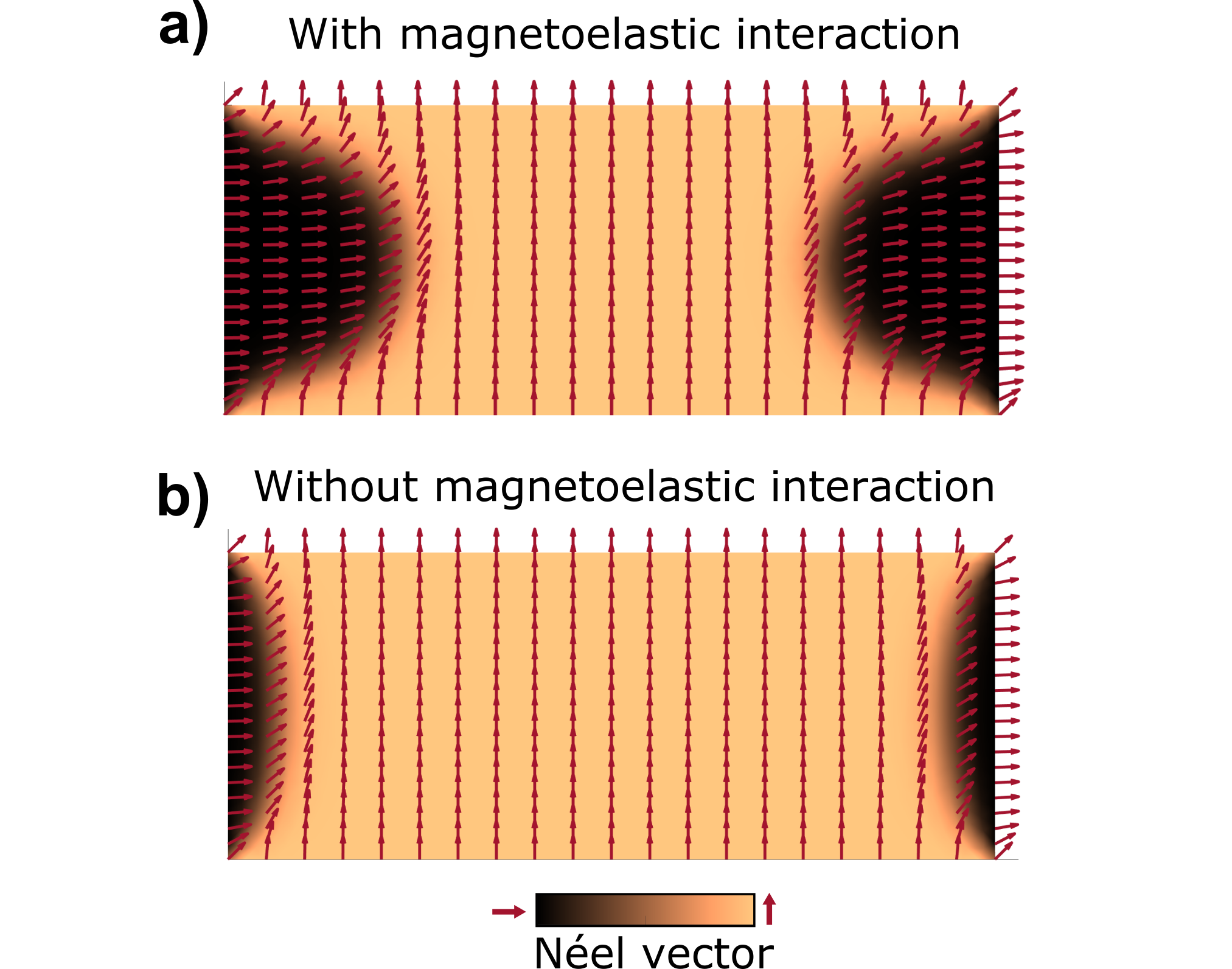}%
\caption{\label{Comparision} Comparison between equilibrium states with (a) and without (b) consideration of magnetoelastic coupling.}
\end{figure}

In this geometry the surface anisotropy induces the formation of the dark domains along $[{\overline{1}}10]$ edges and bright domains along [110] edges. The final texture includes two closure domains localised at the short edges and a large orthogonal domain that spreads between the two long edges. The closure domains grow from the edges due to magnetoelastic forces that act to diminish the average magnetoelastic charge of the sample. This growth is limited by an increase of the energy of the domain walls. The size of the closure domains is of the order of the size of the short edge and depends on the aspect ratio of the sample (see Fig. \ref{fig:3}b). It should be noted that in the absence of magnetoelastic coupling, the closure domains would be localised in the vicinity of the short edge within a distance of several magnetic domain wall widths ( Fig. \ref{Comparision}), independent on the aspect ratio of the device.
Our simulations also show that the closure domains can be localised along the longer edges of the samples as well, as could be experimentally observed for larger patterned devices. However, both configurations (the one with the closure domains along the short and the other along the long edges) are observed in a finite range of aspect ratios (between 1/3 and 3) for which their energies have comparable values. In this case the structure of the final state depends on the initial configuration and kinetics of the domain growth.

\begin{figure}[b]
\includegraphics{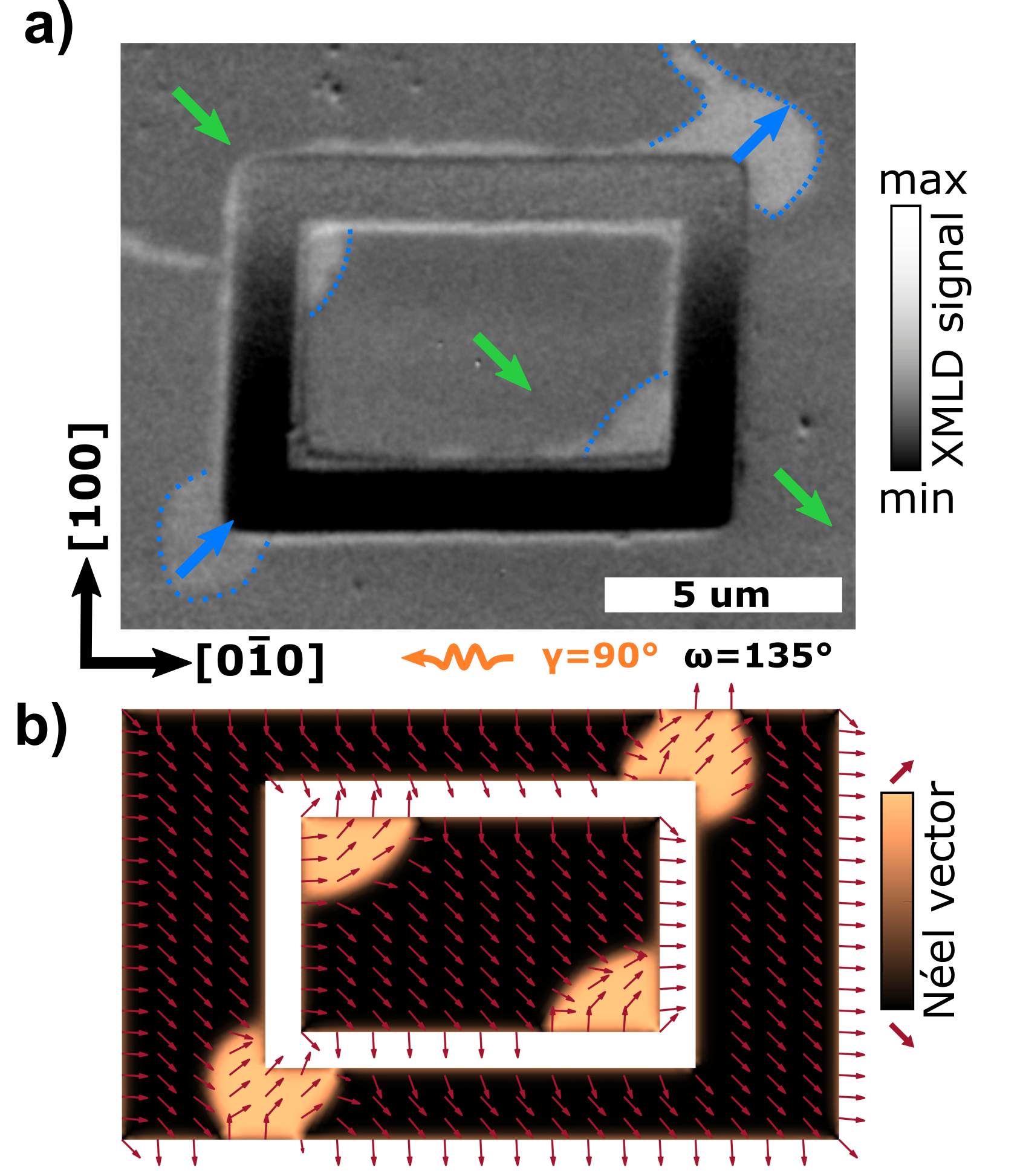}%
\caption{\label{fig:2} (a) Antiferromagnetic domain structure of a rectangular shaped element oriented along the in-plane projections of the hard axes. (b) Simulated equilibrium state of the magnetic texture. The color code and the arrows indicate the direction of the N\'eel vector.}
\end{figure}

To better illustrate the effect of strain, we next investigate elements oriented along the [100] and [010] axes, where even more conspicuous effects are expected. For elements oriented along the projection of the hard axes the domains do not align along the edges of the element, but instead are centered around the corners of the elements, see Fig. \ref{fig:2}a. Inside the element we can observe a large green domain and two blue domains, which are located near the top left and bottom right corner. Outside of the element we observe a domain (green arrows) and two additional domains (blue arrows) located at the top right and bottom left corner, opposite to the domains in the inner corners. In the case that the domain formation is dominated by an alignment along certain crystallographic axes one would expect the same domains to be present at the inside and outside edge of the element. However, this is not the case and we therefore need to consider shape-dependent strain, in particular the role of incompatibility charges in the corners of the elements, to understand the origin of the domain structure.

\begin{figure}
\includegraphics[width=0.5\textwidth]{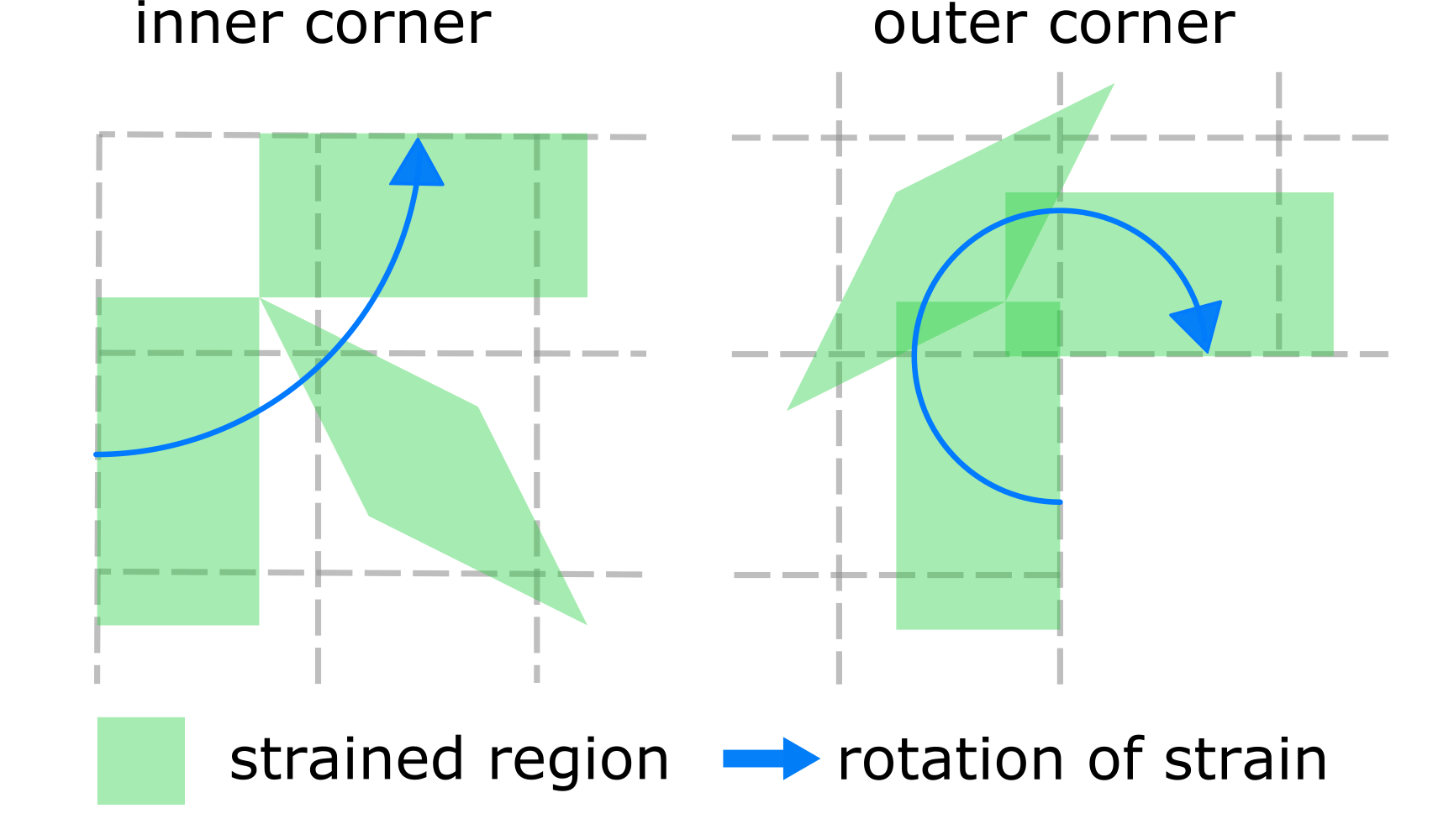}%
\caption{\label{SketchCorner} Sketch of the different direction of the rotation for strain at inner and outer corners.}
\end{figure}

For elements cut along the in-plane projection of the hard magnetic axis, the surface anisotropy favours an orientation of the N\'eel vector along a hard magnetic axis and does not set a preferable domain type. However, the surface anisotropy sets orthogonal easy directions at neighboring edges and favours the formation of vortex-like textures of the N\'eel vectors in the vicinity of the sample corners. Such a rotation of the N\'eel vector through 90$^\circ$ is associated with an inhomogeneous  rotation of the spontaneous strain $\hat{u}^\mathrm{spon}_0$ and creates an elastic vortex structure
 -- so-called disclinations \cite{JohnA.SimmonsR.deWit1970,Kleman1972} -- localized in the corners. Each disclination is characterised by incompatibility charges which have opposite sign in neighboring corners (see Fig. \ref{SketchCorner}). These charges create a radially distributed  field of elastic strain $\hat{u}^\mathrm{elast}$ \cite{Kleman1972} that via magnetoelastic coupling sets preferable directions for the N\'eel vector along the bisectrices of the element. For the elements cut along the easy magnetic axis, this strain couples with the orientation of the N\'eel vector in the center of the domain walls and splits the energy of the domain walls pinned to the neighboring corners. For the elements cut along the hard magnetic axis, it removes the degeneracy between the bright and dark domains and facilitates a formation of the domain of a certain type. In other words, the elastic strains lower the energy barrier for a closure domain.  Here, the closure domain starts to grow from the opposite corners, which have the same sign of rotation, as shown in Fig. \ref{fig:2}(b).
 Interestingly, such magnetoelastic disclinations appear not only at the inner corners of the rectangular elements, but also at the corners of the outer part of the element, where the spontaneous strain rotates in the opposite direction \citep{Kim2018}.

Hence, internal and external strain charges of the same corner have opposite signs. As a result, the closure domains of the same type start to grow along the different diagonals in the internal and external regions (see Fig. \ref{fig:2}(a) and (b)).
Magnetoelastic disclinations (corner charges) are present also in the elements cut along the easy magnetic axes. In this case, corresponding elastic strains set a preferable direction of the N\'eel vector inside the domain walls near the corners. Our calculations show a difference of domain wall width and pinning energy in the neighboring corner as a result of the strain-induced anisotropy.

In addition, different antiferromagnetic domains are accompanied by the deformation of the crystallographic structure. Due to the need for mechanical equilibrium, the creation of antiferromagnetic domains is accompanied by destressing effects \citep{Gomonay2005,Gomonay2007}. The total energy of a patterned element is here complemented by the destressing energy $W_\mathrm{destr}$, which describes the coupling with the nonmagnetic substrate and edge effects, and the surface energy $W_{\text{surf}}$ of the patterned edges. The surface energy is modelled in a way to favor an orientation of the N\'eel vector parallel to the normal $\mathbf{N}$ with respect to the patterned edge:
\begin{equation}\label{eq_surface_energy}
    W_\mathrm{surf}=-K_\mathrm{surf}\oint (\mathbf{n}\cdot\mathbf{N})^2 d\ell.
\end{equation}
Here, the constant $K_\mathrm{surf}>0$ parameterizes the surface energy and $\ell$ is the coordinate along the sample edge. The destressing energy, $W_\mathrm{destr}$, is treated as an additional contribution of the elastic strains $\hat{u}^\mathrm{elast}$, which maintains the strain compatibility of the sample at the interface with nonmagnetic substrate.

\begin{figure}
\includegraphics[width=0.5\textwidth]{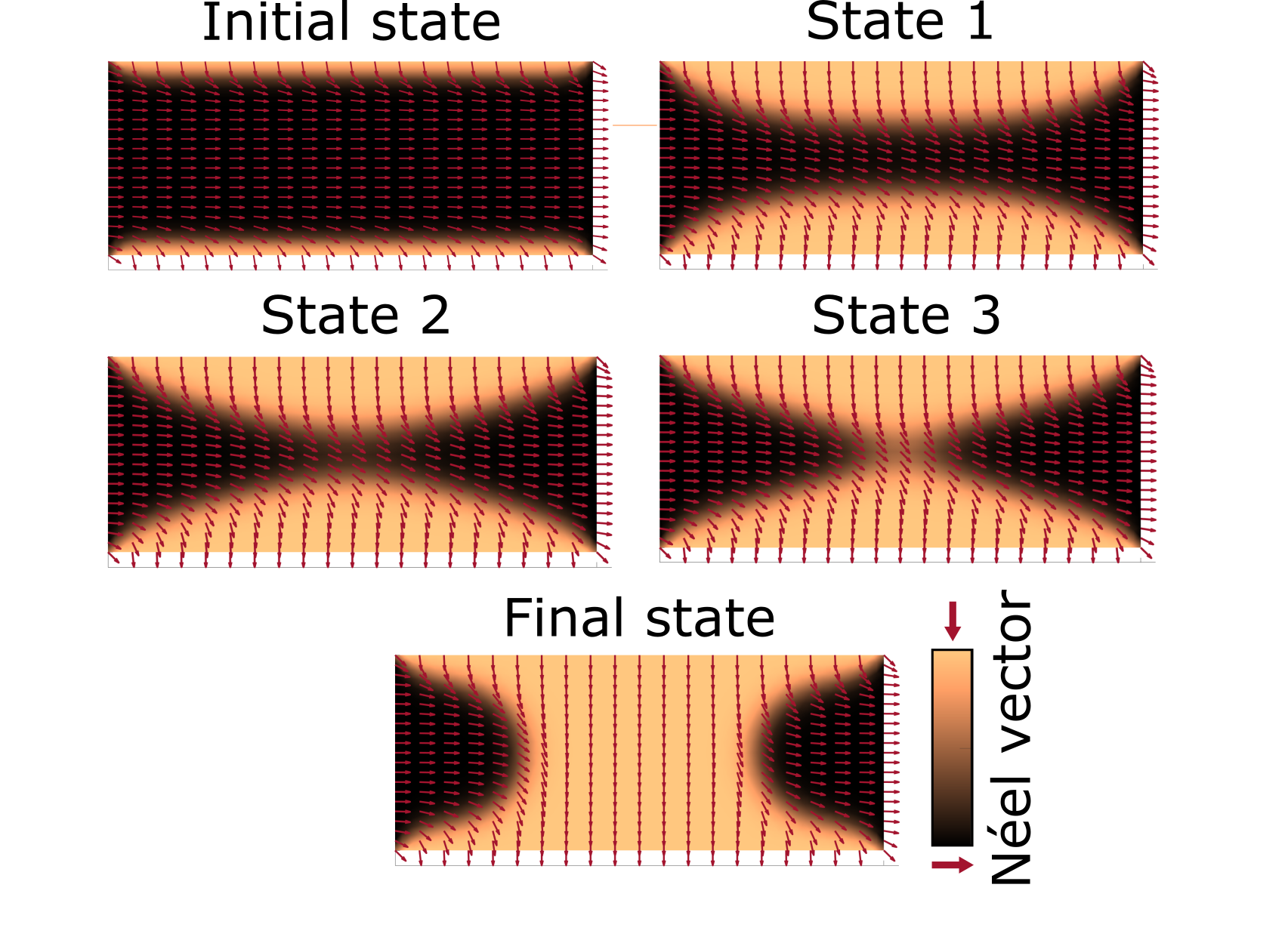}%
\caption{\label{Movie} Simulated evolution of the domain structure from the initial state to the final equilibrium state.}
\end{figure}

To demonstrate the role of incompatibility and the destressing effects in the formation and stabilization of the domain structure, we have calculated the evolution of the domain structure for elements along the in-plane projection of the easy axes, starting from the almost homogeneous state (domain 2, green) with small  domains (domain 1, blue) localised at the long edges of the sample using different values of the damping parameter (different rates of the energy losses). At the initial stage, the closure domains (at the long edge), being pinned in the corners, grow in size trying to reduce the average incompatibility of the sample (see Fig. \ref{Movie}).
In case of slow (quasistatic) relaxation (large damping) the system evolves into a state with the closure domains 1 along the long edges separated by domain 2. In the opposite case of small damping the closure domains merge and the final state corresponds to the closure domains of the type 2 localised  at the short edges.

\section{Discussion}
By investigating elements with different orientations and aspect ratios etched into NiO/Pt bilayers, we identify long-range strain to govern the shape-dependent formation of antiferromagnetic domains.
We observe a preferential orientation of the N\'eel vector perpendicular to the edge of our devices due to patterning effects. Consistent with our previous observations the lattice mismatch between the MgO substrate and the NiO layer leads to a stabilization of different T-domains with the largest out of plane components, due the strain in the out of plane direction \citep{Schmitt2020b}. We investigate the domain structures of elements oriented along the projection of the easy and hard axes and can identify shape-dependent strain to be responsible for the observed domain structures. We can reproduce our experimental observations by magnetoelastic modelling that accounts for the spontaneous strain, due to the distribution of the N\'eel vector, and elastic strain due to contributions from the substrate and the patterning. In previous studies, the substrate-induced strain is the same for the different T-domains \citep{Schmitt2020b}. Here, however, the shear strain due to the domain formation varies for the different domains across the different geometries. The strain that dominates the formation of the domain structure does not arise from patterning induced strain, but from the strain that is associated with the formation of each T-domain. The interactions between these strains are responsible for the formation and stabilization of the equilibrium domain structure.
Analogous to shape-anisotropy in ferromagnets, magnetoelastic interactions in antiferromagnets are long-range and can be used to tailor the antiferromagnetic ground state of antiferromagnetic devices. For example, by choosing the right size, aspect ratio and orientation one could use shape-induced strain to control the antiferromagnetic ground state in antiferromagnetic THz emitters to tailor and optimize their response \citep{Paperfrom2020}. In addition, the strain from the patterned device itself could be used in electrical switching of antiferromagnets to support or hinder the reorientation of the N\'eel vector independent of the underlying switching mechanism.

Finally we note that long range magnetoelastic interactions in AFMs could also be a challenge for the development of antiferromagnetic data storage as reorienting the N\'eel order to switch bits will by the change in strain affect neighbouring bits. However, by considering physical separation, as in established bit-patterned media \citep{White1997}, the non interacting nature of AFMs can be fully taken advantage of due to the absence of magnetic stray fields.

In summary, we identify how shape-dependent strain can be used to control the antiferromagnetic ground state in NiO over several microns. Since magnetoelastic coupling is significant for several other antiferromagnets such as CoO and Hematite, shape-induced strain can be considered to be the antiferromagnetic equivalent of conventional shape-induced anisotropy in ferromagnets and provide a unique means to control antiferromagnets.

\begin{acknowledgments}

The authors thank T. Reimer for skillful technical assistance. We thank HZB for the allocation of synchrotron radiation beamtime, we thankfully acknowledge the financial support by HZB. The work has benefited from insights gained from experiments that were performed at the CIRCE beamline at ALBA Synchrotron with the collaboration of ALBA staff. L.B. acknowledges the European Union’s Horizon 2020 research and innovation program under the Marie Skłodowska-Curie Grant Agreement ARTES No. 793159. L.B. and M.K. acknowledge support from the Graduate School of Excellence Materials Science in Mainz (MAINZ) DFG 266, the DAAD (Spintronics network, Project No. 57334897 and Insulator Spin-Orbitronics, Project No. 57524834), and all groups from Mainz acknowledge that this work was funded by the Deutsche Forschungsgemeinschaft (DFG, German Research Foundation), TRR 173-268565370 (Project Nos. A01, A03, A11, B02, and B12) and KAUST (OSR-2019-CRG8-4048). J.S. additionally acknowledges the Alexander von Humboldt Foundation and O.G and J.S. acknowledge the EU FET Open RIA Grant No. 766566 and the Deutsche Forschungsgemeinschaft (DFG, German Research Foundation), TRR 288-422213477 (project A09) and the Grant Agency of the Czech Republic grant no. 19-28375X. R.R. also acknowledges support from the European Commission through the Project 734187-SPICOLOST (H2020-MSCA-RISE-2016), the European Union’s Horizon 2020 research and innovation program through the Marie Sklodowska-Curie Actions Grant Agreement SPEC No. 894006, the MCIN/AEI (RYC 2019-026915-I), the Xunta de Galicia (ED431B 2021/013, Centro Singular de Investigación de Galicia Accreditation 2019-2022, ED431G 2019/03) and the European Union (European Regional Development Fund - ERDF). M.K. acknowledges financial support from the Horizon 2020 Framework Programme of the European Commission under FET-Open Grant Agreement No. 863155 (s-Nebula) and from the Research Council of Norway through its Centers of Excellence funding scheme, project number 262633 “QuSpin”. This work was also supported by ERATO “Spin Quantum Rectification Project” (Grant No. JPMJER1402) and the Grant-in-Aid for Scientific Research on Innovative Area, “Nano Spin Conversion Science” (Grant No. JP26103005), Grant-in-Aid for Scientific Research (S) (Grant No. JP19H05600) from JSPS KAKENHI, Japan.

\end{acknowledgments}

\appendix

\section{Sample preparation}
Similar to previous studies \citep{Schmitt2020b,Baldrati2019} we used reactive magnetron sputtering to grow epitaxial MgO(001)//NiO(10nm)/Pt(2nm) thin films. Before deposition the MgO(001) substrates were pre-annealed at $770^\circ$ for $2\,$h in vacuum. We then deposited NiO from a Ni target at $430^\circ$C and $150\,$W in an atmosphere of O$_2$ (flow 3$\,$sccm) and Ar (flow 15$\,$sccm), before depositing the platinum layer in situ at room temperature.

\section{XMLD signal}\label{sec_XMLDsignal}

\begin{figure}
\includegraphics[width=0.47\textwidth]{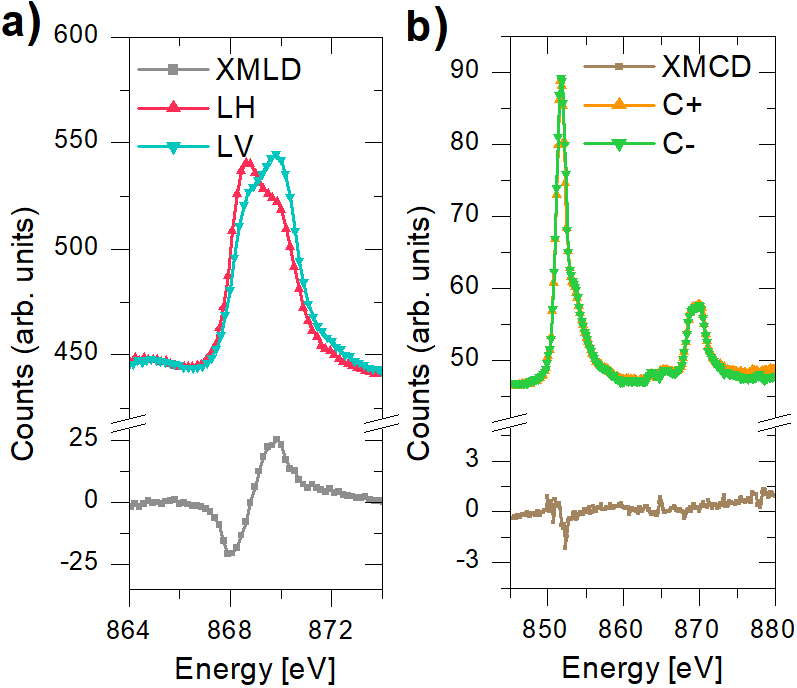}%
\caption{\label{Sup1} (a) X-ray absorption spectrum and XMLD for linear polarized X-rays in the vicinity of the L$_2$ edge. (b) X-ray absorption spectrum and XMCD for circular polarized X-rays.}
\end{figure}

We observed X-ray magnetic linear dichroism (XMLD) in Fig. \ref{Sup1}(a), calculated as I$_{\text{LH}}$-I$_{\text{LV}}$, and an absence of circular magnetic dichroism (XMCD) in Fig. \ref{Sup1} (b), indicating purely antiferromagnetic ordering of our films. For subsequent XMLD imaging we utilized the inversion of the XMLD contrast at the Ni L$_2$ edge at the energies $E_1$=868.3$\,$eV, $E_2$=870.1$\,$eV and calculated the XMLD image by [$I(E_1$)-$I(E_2$)]/[$I(E_1$)+$I(E_2$)]. By studying the XMLD contrast dependence on the azimuthal angle $\gamma$ and angle of the beam-polarization $\omega$ we can identify in Fig. \ref{fig:1}(a) four different antiferromagnetic domains to be present in a rectangular element of $10\times5\,$\textmu m whose long axis is oriented along the [110] direction.

\begin{figure}
\includegraphics[width=0.5\textwidth]{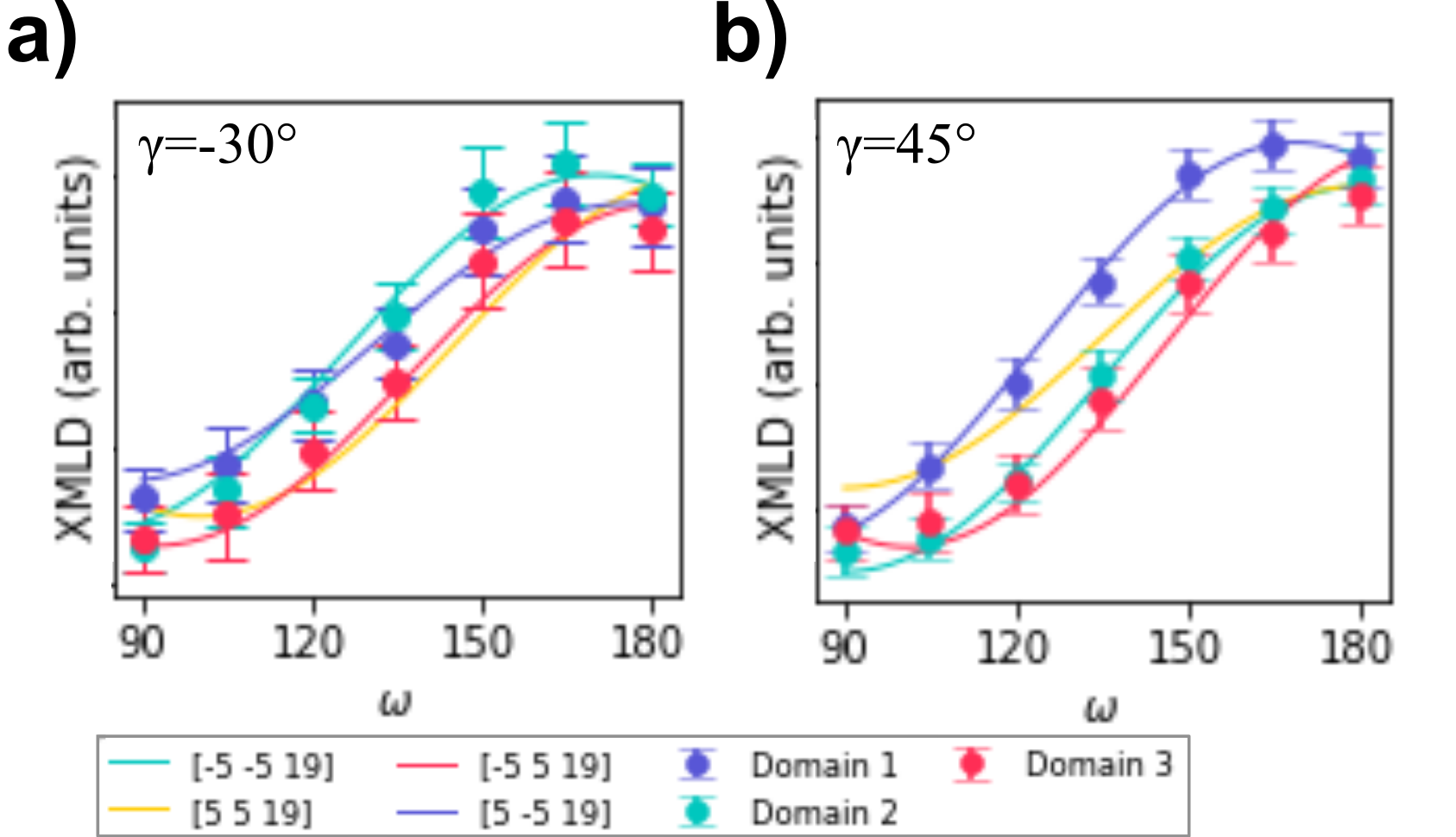}%
\caption{\label{Sup2} Simulated and experimental contrast of the different domains dependent on the beampolarization for (a) $\gamma=-30^\circ$ and (b) $\gamma=45^\circ$.}
\end{figure}

NiO is a compensated collinear antiferromagnet with two magnetic sublattices whose magnetic structure is uniquely described by the N\'eel vector $\mathbf{n}$. In bulk NiO crystals the exchange-striction leads to a rhombohedral contraction along the $\langle 111 \rangle $ axes. Thus, four twin T-domains can be formed with the spins being ferromagnetically coupled inside the $\{111\}$ planes and antiferromagnetically between the $\{111\}$ planes, due to the superexchange interaction. Within such a T-domain the spins can be oriented along three possible easy axes $\langle 11\bar{2} \rangle$, leading to three different spin domains S-domains per T-domain. There are a total of 12 possible domains in bulk NiO crystals \citep{Roth1960, Roth1960a, Slack1960}. However, in NiO thin films the growth induced strain from the substrate mismatch can lead to a preferential stabilization of S-domains parallel or perpendicular to the sample plane \citep{Alders1998, Altieri2003}.
In a previous study on similarly grown MgO//NiO(10)/Pt(2) thin films we observed that the epitaxial growth of NiO on MgO substrates results in compressive strain, which lead to a preferential out of plane component of the N\'eel  vector \citep{Schmitt2020b, Alders1998, Altieri2003}. Here, we have observed analogous contrast changes among the domains with varying azimuthal angle $\gamma$ and orientation of the linear beam polarization $\omega$ in Fig. \ref{Sup2}. Thus, we can identify the orientation of the N\'eel vector of our domains to be along the $[ \pm 5  \pm 5$ $19]$ directions, in line with our previous findings \citep{Schmitt2020b}.


\section{Modelling of the magnetic textures}\label{sec_theory}
We model these contributions according to the tetragonal crystallographic symmetry of the NiO film (deformed cubic crystal due to the substrate), so that the density of the magnetic energy is given by4
\begin{equation}\label{eq_magnetic}
    w_\mathrm{mag}=\frac{1}{2}A\left(\nabla\varphi\right)^2- \frac{1}{4}H_\mathrm{an}M_{\text{s}}\cos4\varphi,
\end{equation}
 where $A$ is the exchange stiffness of NiO, $H_\mathrm{an}$ is the anisotropy field, and $M_{\text{s}}/2$ is the saturation magnetization of the magnetic sublattice.
 The densities of magnetoelastic and elastic energy are
 \begin{widetext}
 \begin{eqnarray}\label{eq_magnetoelastic}
    w_\mathrm{m-e}&=&\lambda_\mathrm{me} M^2_s \left[\cos2\varphi(u_{xx}-u_{yy})+2\sin2\varphi u_{xy}\right],\\
    w_\mathrm{elas}&=&\frac{1}{2}\mu\left[(u_{xx}-u_{yy})^2+4u^2_{xy}\right]+\frac{\mu}{2(1-2\nu)}(u_{xx}+u_{yy})^2,\nonumber
\end{eqnarray}
\end{widetext}
 where $\lambda_\mathrm{me}$ is the magnetoleastic parameter, $\mu$ is the shear modulus, $\nu$ is the Poisson ratio. In Eq.~(\ref{eq_magnetoelastic}), we omit the out-of plane components of strain ($u_{zz}$, $u_{xz}$, $u_{yz}$) and disregard the coupling between the N\'eel vector and the isotropic strain ($u_{xx}+u_{yy}$), since these  effects are irrelevant for the formation of the domain structure. The coordinate axes $x$ and $y$  are aligned along the easy magnetic axis $[1\overline{1}0]$ and $[110]$. General expressions for these contributions can be found in \cite{Schmitt2020b, Gomonay2021}.

The  strain in the NiO layer is represented as a sum of spontaneous (plastic) strains and elastic strains, as described in the main part: $\hat{u}^\mathrm{tot}=\hat{u}^\mathrm{spon}_0+\hat{u}^\mathrm{elast}$. The spontaneous strains are calculated by minimizing the elastic and magnetoelastic energy contributions (\ref{eq_magnetoelastic}) for a given value of $\varphi(x,y)$:
\begin{eqnarray}\label{eq_spontaneous}
    u^\mathrm{spon}_{xx}&=&-u^\mathrm{spon}_{yy}=-\frac{\lambda_\mathrm{me}}{\mu}M^2_{\text{s}}\cos2\varphi,\nonumber\\ u^\mathrm{spon}_{xy}&=&-\frac{\lambda_\mathrm{me}}{\mu}M^2_{\text{s}}\sin2\varphi.
\end{eqnarray}

The elastic strains $\hat{u}^\mathrm{elast}$ are then calculated from the compatibility equations
\begin{equation}\label{eq_incompatibility_A}
\varepsilon_{ijk}\varepsilon_{lmn}\partial_j\partial_m{u}^\mathrm{elast}_{kn}  =\eta_{il},
\end{equation}
The incompatibility arising at the interface between the magnetic and the nonmagnetic layer is calculated as
\begin{eqnarray}\label{eq_incompatilibity_interface}
    \eta_{xx}&=&-\eta_{yy}=\frac{\lambda_\mathrm{me}}{\mu}(\partial_z M_{\text{s}})^2\cos2\varphi,\nonumber\\    \eta_{xy}&=&\frac{\lambda_\mathrm{me}}{\mu}(\partial_z M_{\text{s}})^2\sin2\varphi.
\end{eqnarray}
The magnetic order, parameterized as $M_{\text{s}}(z)$, vanishes at the scale of the exchange length $\xi_\mathrm{exch}$ ($\propto$ several inter atomic distances) , which is much smaller than the other length scales in the system (domain wall width, thickness of the NiO layer, etc). We can assume that the incompatibility charges (\ref{eq_incompatilibity_interface}) are localized at the interface $z=0$ and substitute $(\partial_z M_{\text{s}})^2\Rightarrow M_{\text{s}}^2\delta(z)/\xi_\mathrm{exch}$.
For the elastic strains $\hat{u}^\mathrm{elast}$ that are created by incompatibility charges (\ref{eq_incompatilibity_interface}), we use the Eshelby solution \cite{Eshelby1956}, so that
\begin{eqnarray}\label{eq_Eshelby_solution}
    u^\mathrm{elas}_{xx}(\mathbf{r})&=&-u^\mathrm{elas}_{yy}(\mathbf{r})=\frac{u_0}{4\pi}\int d\mathbf{r}\prime\frac{\cos2\varphi(\mathbf{r}^\prime)}{|\mathbf{r}-\mathbf{r}^\prime|},\nonumber\\
    u^\mathrm{elas}_{xy}(\mathbf{r})&=&\frac{u_0}{4\pi}\int d\mathbf{r}\prime\frac{\sin2\varphi(\mathbf{r}^\prime)}{|\mathbf{r}-\mathbf{r}^\prime|},
\end{eqnarray}
where $u_0\equiv\lambda_\mathrm{me}M^2_s/\mu$ is the value of the spontaneous strain.

By substituting Eqs.~(\ref{eq_Eshelby_solution}) into Eq.~(\ref{eq_general_energy}) we get the expression for the destressing energy as
\begin{equation}\label{eq_destressing}
    W_\mathrm{destr}=\kappa \int d\mathbf{r}\int d\mathbf{r}^\prime\frac{\cos2\left[\varphi(\mathbf{r}^\prime)-\varphi(\mathbf{r})\right]}{|\mathbf{r}-\mathbf{r}^\prime|}.
\end{equation}
Here, $\kappa$ is a phenomenological coefficient that generally depends on the elastic properties of both the magnetic and nonmagnetic layers and the interface. In a simplified model with isotropic magnetoelastic and elastic interactions $\kappa\propto \mu u_0^2/(4\pi)$.
We calculate the distribution of the magnetic variable $\varphi(\mathbf{r})$ from the following set of equations, which minimizes the  energy of the sample $W_\mathrm{bulk}+W_\mathrm{destr}$:
\begin{widetext}
\begin{eqnarray}\label{eq_set_magnetoelastic}
    -x^2_\mathrm{DW}\Delta \varphi+\sin 4\varphi=\frac{2\kappa}{H_\mathrm{an}M_{\text{s}}}\int d\mathbf{r}^\prime\frac{\sin2\left[\varphi(\mathbf{r}^\prime)-\varphi(\mathbf{r})\right]}{|\mathbf{r}-\mathbf{r}^\prime|}
    &+&\frac{2\lambda_\mathrm{me}M_{\text{s}}}{H_\mathrm{an}}\left[(u_{xx}-u_{yy})\sin2\varphi-2 u_{xy}\cos2\varphi\right],\nonumber\\
    \Delta \mathbf{u}+\frac{1}{1-2\nu} \nabla(\nabla\cdot\mathbf{u})&=&2u_0\left(\begin{array}{c}
        \partial_x(\cos2\varphi)+  \partial_y(\sin2\varphi) \\
          \partial_x(\sin2\varphi)-  \partial_y(\cos2\varphi)
    \end{array}\right)
\end{eqnarray}
\end{widetext}
where $x_\mathrm{DW}\equiv\sqrt{A/H_\mathrm{an}M_{\text{s}}}$ is the  domain wall width, and $\mathbf{u}$ is the displacement vector. The second and third equations correspond to conditions of mechanical equilibrium (zero stresses) inside the NiO layer.

Equations (\ref{eq_set_magnetoelastic}) are complemented by the boundary conditions at the edges:
\begin{widetext}
\begin{eqnarray}\label{eq_boundary_conditions}
    &-&A\mathbf{N}\cdot\nabla\varphi+K_\mathrm{surf}\left[(N_x^2-N_y^2)\sin2\varphi+2N_xN_y\cos2\varphi\right]=0,\nonumber\\
    &-&\mathbf{N}\cdot\nabla \mathbf{u}+2u_0\left(\begin{array}{c}
        N_x\cos2\varphi+N_y\sin2\varphi \\
          N_x\sin2\varphi- N_y\cos2\varphi
    \end{array}\right)=0.
\end{eqnarray}
\end{widetext}
While the strain distribution within the film plane can be calculated directly from the elasticity equations (\ref{eq_set_magnetoelastic}), we also introduce an approach based on the associated incompatibility component
\begin{equation}\label{eq_incompatibilityzz}
    \eta_{zz}\equiv -u_0\left[(\partial_x^2-\partial_y^2)\cos 2\varphi-2\partial_x\partial_y\sin2\varphi\right].
\end{equation}
 This component is mainly concentrated in the sample corners, where it creates partial disclinations and facilitates the formation of domains. To understand this effect, we consider the immediate vicinity of a corner. Strong surface anisotropy aligns the N\'eel vector at the edges along $\mathbf{N}$ direction. Inside the sample  the N\'eel vector rotates through 90$^\circ$ to fit the boundary conditions. Due to the magnetoelastic mechanism, such a rotation creates a nonzero lattice  curvature, which is described by the components of the bend-twist tensor \cite{Kleman1972}
 \begin{equation}\label{eq_bend_twist_tensor}
     K_{zx}=-\frac{u_0 Y}{\sqrt{X^2+Y^2}},\quad  K_{zy}=\frac{u_0 X}{\sqrt{X^2+Y^2}},
 \end{equation}
 with the $XY$ coordinate frame centered at the corner.
 According to Eq.~(\ref{eq_incompatibilityzz}) the associated incompatibility $\eta_{zz}\approx\pm 2u_0/(X^2+Y^2)$. The sign of $\eta_{zz}$ depends on the direction of the rotation and is opposite for the neighboring corners, see Fig.\ref{SketchCorner}.

 Such a distribution of lattice curvature and incompatibility charge can be treated as an elastic defect, known as a straight disclination line \cite{DeWit1973}, with the Frank vector (vector of disclination) $\Omega=2\pi u_0$ pointing along $z$ direction. Similar  disclinations of magnetoelastic origin that appear at the junctions of several domain walls in ferromagnets were described by Kleman et al. \cite{Kleman1972}. The elastic strain created by the disclination according to \cite{Kleman1972} are
\begin{eqnarray}\label{eq_strain_field_disclination}
  u^\mathrm{elas}_{xx}&-&u^\mathrm{elas}_{yy}= -\frac{u_0}{2(1-\nu)}\frac{X^2-Y^2}{X^2+Y^2},\nonumber\\ u^\mathrm{elas}_{xy}&=& -\frac{u_0}{2(1-\nu)}\frac{XY}{X^2+Y^2}.
\end{eqnarray}
The strain component $u^\mathrm{elas}_{xy}$, which is maximal along bisectrix $X=Y$, sets the preferable direction of the N\'eel vector either along or perpendicular to the bisectrix direction, depending on the sign of the incompatibility charge. It should be mentioned that incompatibility charges in the inner and outer corners have opposite signs (due to opposite direction of the rotation), thus favoring the formation of domains of different types.

Numerical simulations of the textures were implemented by solving Eqs.~(\ref{eq_set_magnetoelastic}), (\ref{eq_boundary_conditions}) with PDE Tools of Matlab2021 with the following parameters: $x_\mathrm{DW}=100$~nm, $u_0=3\cdot 10^{-5}$, $H_\mathrm{an}=0.03$~T, $M_s=10^6$A/m,  $\mu=35$~GPa. The equilibrium structure was calculated by the relaxation from different initial states with the loops for self-correction of the destressing energy, see Fig. \ref{Movie}.
To obtain a multidomain structure, we always introduced a domain wall into the system (either 180$^\circ$ or 90$^\circ$).
\providecommand{\noopsort}[1]{}\providecommand{\singleletter}[1]{#1}%

\end{document}